\documentclass[aip,apl,twocolumn,12pt,reprint]{revtex4-1}
\usepackage{graphicx}
\usepackage{amsmath}
\begin{document}

\title{Spin-orbit torque opposing the Oersted torque in ultrathin Co/Pt bilayers}
\author{T.D. Skinner}
\email[Electronic mail: ]{tds32@cam.ac.uk}
\affiliation{Cavendish Laboratory, University of Cambridge, CB3 0HE, United Kingdom}

\author{M. Wang}
\affiliation{School of Physics and Astronomy, University of Nottingham, NG7 2RD, United Kingdom}

\author{A.T. Hindmarch}
\altaffiliation[Current address: ]{Centre for Materials Physics, Durham University, DH1 3LE, United Kingdom}
\affiliation{School of Physics and Astronomy, University of Nottingham, NG7 2RD, United Kingdom}

\author{A.W. Rushforth}
\affiliation{School of Physics and Astronomy, University of Nottingham, NG7 2RD, United Kingdom}

\author{A.C. Irvine}
\affiliation{Cavendish Laboratory, University of Cambridge, CB3 0HE, United Kingdom}

\author{D. Heiss}
\altaffiliation[Current address: ]{COBRA Research Institute, Eindhoven University of Technology, Postbus 513, 5600 MB Eindhoven, The Netherlands}
\affiliation{Cavendish Laboratory, University of Cambridge, CB3 0HE, United Kingdom}

\author{H. Kurebayashi}
\altaffiliation[Current address: ]{London Centre for Nanotechnology, University College London, WC1H 0AH, United Kingdom}
\affiliation{Cavendish Laboratory, University of Cambridge, CB3 0HE, United Kingdom}

\author{A.J. Ferguson}
\email[Electronic mail: ]{ajf1006@cam.ac.uk}
\affiliation{Cavendish Laboratory, University of Cambridge, CB3 0HE, United Kingdom}

\date{\today}
\begin{abstract}
Current-induced torques in ultrathin Co/Pt bilayers were investigated using an electrically driven FMR technique. The angle dependence of the resonances, detected by a rectification effect as a voltage, were analysed to determine the symmetries and relative magnitudes of the spin-orbit torques. Both anti-damping (Slonczewski) and field-like torques were observed. As the ferromagnet thickness was reduced from 3 to 1 nm, the sign of the field-like torque reversed. This observation is consistent with the emergence of a Rashba spin orbit torque in ultra-thin bilayers. 
\end{abstract}
\maketitle

\begin{figure}
\centering
\includegraphics[]{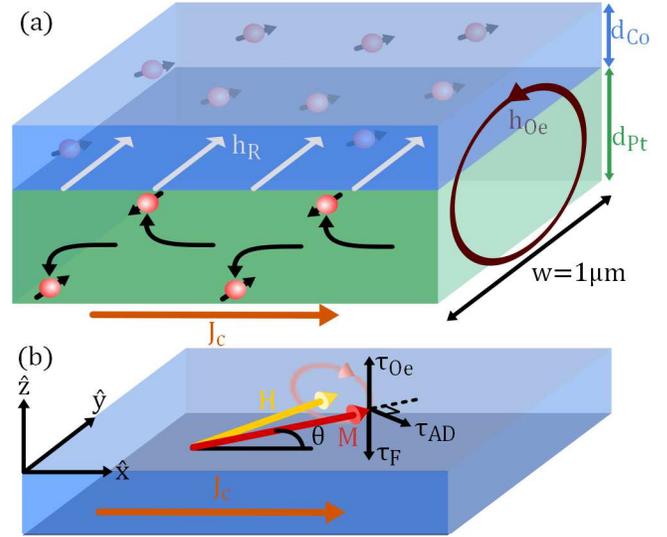}
\caption{(a) A charge current density, $J_{\mathrm{C}}$ passing through the bilayer induces a transverse spin-current in the platinum due to the spin-Hall effect which flows into the cobalt layer. At the interface, due to the structural inversion asymmetry, the conduction electrons experience an effective magnetic field, $h_{\mathrm{R}}$. The cobalt has an additional oxidised silicon interface which could also similarly produce an effective magnetic field. The current passing through the platinum also induces an Oersted field in the cobalt, due to Amp\`{e}re's law. For clarity, the bilayers  drawn here are inverted from our actual structure. (b) The Oersted field induces an out of plane torque on the cobalt magnetisation, $\tau_{\mathrm{Oe}}$. Additional anti-damping and field-like torques, $\tau_{\mathrm{AD}}$ and $\tau_{\mathrm{F}}$ respectively, are induced due to the exchange interaction of the non-equilibrium spin-density in the ferromagnet with the magnetisation. A field-like torque with negative coefficient is shown here opposing a positive Oersted torque.}	
\label{im:device}
\end{figure}

Current-induced spin-orbit torques in ultrathin ferromagnetic/heavy metal bilayers provide ways to electrically control magnetisation. Two mechanisms for observed torques have been proposed, both of which could contribute to the total torques and both of which originate in the spin-orbit interaction. A schematic of both mechanisms is shown in figure \ref{im:device}a. The first mechanism is due to the spin-Hall effect,\cite{hirsch1999spin, sinova2004universal, kato2004observation, wunderlich2005experimental} where a charge-current in the heavy metal layer generates spin currents perpendicular to the charge-current. When a spin-current flows into the ferromagnetic layer, it can exert a spin-transfer torque (STT).\cite{Liu2011,Liu04052012,Liu2012current} This torque normally follows the anti-damping form predicted by Slonczewski\cite{slonczewski1996current} and Berger,\cite{berger1996emission} but it is known that a field-like non-adiabatic spin transfer torque can also exist.\cite{Zhang2002,zimmler2004,sankey2007measurement} 

The second mechanism is a `Rashba' spin-orbit torque. Due to the structural inversion asymmetry of the two dissimilar materials at the interface, when a current is applied, the spin-orbit Hamiltonian breaks the degeneracy of the electron spin states near the interface, creating a non-equilibrium spin-accumulation. The electron spins in the ferromagnet, through exchange coupling, can then exert a torque on the magnetic moments. This was initially predicted to give a field-like torque, acting perpendicularly to the interface normal and injected current,\cite{Manchon2008, Obata2008, Manchon2009} which was later confirmed by experiments in ultrathin Pt/Co/AlO$_x$ \cite{miron2010current, miron2011fast, pi2010tilting} and Ta/CoFeB/MgO \cite{suzuki2011current} trilayers. However, further measurements in these layers have confirmed the presence of an additional anti-damping torque.\cite{miron2011perpendicular,Garello2013} A recent experiment, in a single-layer ferromagnet with broken symmetry, has shown that this anti-damping torque can be explained by the precession of the spins, initially polarised along the magnetisation, around the additional current-induced spin-orbit fields.\cite{kurebayashi2013observation} These additional torques have also been modelled theoretically in metal bilayer systems.\cite{Pesin2012,Wang2012}

The torques are further complicated by the additional Oersted torque in the ferromagnetic layer, due to the total current in the heavy metal, which has the same symmetry as the field-like torque. The total torques can be formulated as
\begin{equation}
\boldsymbol{\tau} = \tau_{\mathrm{AD}}\mathbf{\hat{m}\times\hat{y}\times\hat{m}}-\left(\tau_{\mathrm{F}}+\tau_{\mathrm{Oe}}\right)\mathbf{\hat{y}\times\hat{m}},
\end{equation}
where the anti-damping ($\tau_{\mathrm{AD}}$) and field-like ($\tau_{\mathrm{F}}$) torques can have contributions from both the spin-Hall and Rashba effects. Previous studies have tried to disentangle these two effects by studying the dependence of the torques on the thickness of the two layers.\cite{kim2012layer,fan2013observation} In particular, Fan \textit{et al.} observed an additional field-like torque in Py/Pt layers with the same direction as the Oersted field.\cite{fan2013observation} In this paper, we report a similar field-like torque, emerging only in the ultra-thin Co layer regime, opposing the Oersted field. This suggests that the field-like torque is sensitive to details of the sample composition and growth and can vary significantly, possibly due to competing mechanisms.

\begin{figure}
\centering
\includegraphics[]{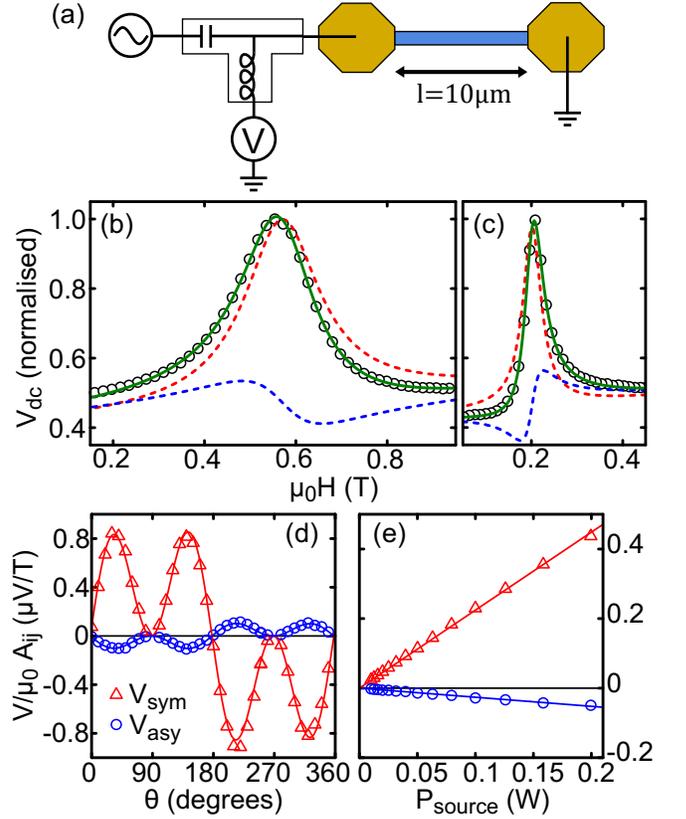}
\caption{(a) To perform the FMR measurement, a bias tee is used to separate the microwave power used to drive the FMR from the dc voltage in which the resonance is detected. (b) The resonance in $V_{\mathrm{dc}}$ is fitted by a combination of antisymmetric and symmetric Lorentzians. In this $d_{\mathrm{Co}}=1$ nm device, the antisymmetric part is negative in amplitude. (c) In devices with $d_{\mathrm{Co}}=3$ nm, the sign of the antisymmetric part becomes positive, whilst the symmetric part remains positive. (d) The angular dependence of $V_{\mathrm{sym}}/A_{yz}$ and $V_{\mathrm{asy}}/A_{yy}$ (shown for a device with $d_{\mathrm{Co}}=1.5$ nm) are fitted well by an in-plane anti-damping torque ($\tau_{\mathrm{AD}}$) and an out of plane field-like driving torque ($\tau_{\mathrm{Oe}}+\tau_{\mathrm{F}}$) respectively. (e) Both voltages peaks observed scale linearly with the microwave source power, as expected from the theoretically linear dependence on current of the spin-Hall and Rashba effects.}	
\label{im:lorentz}
\end{figure}

Using electrically driven FMR,\cite{Liu2011,fang2011spin} we have studied sputtered ultrathin bilayers of Co/Pt which are in-plane magnetised, where the cobalt thickness, $d_{\mathrm{Co}}$, is varied between 1 and 3 nm, whilst the platinum thickness, $d_{\mathrm{Pt}} = 3$ nm, remains constant.  A schematic of the magnetisation precession, and the directions of the torques in our measurement is shown in figure \ref{im:device}b. The layers were deposited on a thermally oxidised silicon substrate by dc magnetron sputtering. 1 x 10 $\mu$m bars were patterned using electron-beam lithography and Ar ion-milling. Each bar was mounted on a low-loss dielectric circuit board. Microwave power was delivered to the board via a semi-rigid coaxial cable. This was connected to a microstrip transmission line on the circuit board which was terminated by a wirebond to one end of the sample. The other end of the sample was connected to ground via another wirebond. An on-board bias-tee,\cite{fang2012electrical} comprising of an in-line gap capacitor and a wirebond as an inductor, was used to separate the injected microwave power from the measurement of the dc voltage, V$_{\mathrm{dc}}$, across the bar (see figure \ref{im:lorentz}a).

The microwave current injected into the bar, I$_0e^{j\omega t}$, induces effective magnetic fields, $(h_{\mathrm{x}},h_{\mathrm{y}},h_{\mathrm{z}})e^{j\omega t}$, which drive FMR. As the magnetisation precesses, there is an oscillating component of the resistance due to the anisotropic magnetoresistance (AMR) of the sample: $R = R_0 + \Delta R\cos^2\theta$, where $\theta$ is the angular separation of the current and magnetisation. At resonance, this rectifies with the driving microwave current to give a peak in V$_{\mathrm{dc}}$. This can be fitted by a combination of symmetric and antisymmetric Lorentzians,\cite{fang2011spin}
\begin{multline}\label{eq:V_dc}
V_{\mathrm{dc}} = V_{\mathrm{sym}}\dfrac{\Delta H^2}{(H - H_0)^2 +\Delta H^2}\\
+V_{\mathrm{asy}}\dfrac{(H - H_0)\Delta H}{(H - H_0)^2 +\Delta H^2},
\end{multline}
where $V_{\mathrm{sym}}$ and $V_{\mathrm{asy}}$ are given by
\begin{equation}
V_{\mathrm{sym}} = V_{\mathrm{mix}}A_{yz}h_{\mathrm{z}}\sin 2\theta
\end{equation}
and
\begin{equation}
V_{\mathrm{asy}} = V_{\mathrm{mix}}A_{yy}(h_{\mathrm{y}}\cos\theta-h_{\mathrm{x}}\sin\theta)\sin 2\theta.
\end{equation}

In these expressions, $V_{\mathrm{mix}}=\frac{1}{2}I_0\Delta R$, $H_0$ and $\Delta H$ are the resonant field and linewidth and $A_{yz}$ and $A_{yy}$ are related to the scalar amplitudes of the ac magnetic susceptibility by $A_{ij} = \chi_{ij}/M_{\mathrm{S}}$. Their values are
\begin{equation}\label{eq:Ayz}
A_{yz}=\dfrac{\sqrt{H_0(H_0+M_{\mathrm{eff}})}}{\Delta H(2H_0 + M_{\mathrm{eff}})},
\end{equation}
and
\begin{equation}\label{eq:Ayy}
A_{yy}=\dfrac{(H_0+M_{\mathrm{eff}})}{\Delta H(2H_0 + M_{\mathrm{eff}})}.
\end{equation}
$M_{\mathrm{eff}}$ is the effective magnetisation which contains a uniaxial interface anisotropy term,
\begin{equation}\label{eq:Meff}
M_{\mathrm{eff}} = M_{\mathrm{S}} - \dfrac{H_{\mathrm{U}}^{\mathrm{int}}}{d_{\mathrm{Co}}},
\end{equation}
which depends on the thickness of the cobalt layer, $d_{\mathrm{Co}}$. We do not find voltages with a symmetry of $\sin\theta$, indicating a negligible spin pumping signal. This is because the rectifying microwave current is much larger than the direct current induced via the inverse SHE acting on spin current pumped by the induced precession.

Source microwave powers of 20 dBm were typically used to excite FMR. Microwave frequencies of between 16 and 19 GHz were used to ensure that the entire resonance peak was measured in a magnetic field large enough that the magnetisation was saturated. With microwave power applied, the resonances were measured in V$_{\mathrm{dc}}$ as the external magnetic field, H, was swept from high to low at an in-plane angle, $\theta$. The resonances were measured for successive values of $\theta$, with the peaks then fitted by equation \ref{eq:V_dc} (figures \ref{im:lorentz}b and c). By measuring FMR out of plane and self-consistently fitting the Kittel and energy equations, values of $M_{\mathrm{eff}}$ were determined.\cite{Ando2011} The fitted values of $\mu_0M_{\mathrm{eff}}$ were similar to those we have previously reported in spin-pumping measurements of the same layers at 250 K,\cite{skinner2013} in this case varying from 1.4 to 0.14 T as $d_{\mathrm{Co}}$ is reduced, consistent with equation \ref{eq:Meff}. To analyse the data, $A_{yz}$ and $A_{yy}$ are calculated from equations \ref{eq:Ayz} and \ref{eq:Ayy} using the measured values of $M_{\mathrm{eff}}$. 

For all the sweeps measured, the symmetric part dominates the antisymmetric part. We now fit the effective fields to $V_{\mathrm{sym}}/A_{\mathrm{yz}}$ and $V_{\mathrm{asy}}/A_{\mathrm{yy}}$ (figure \ref{im:lorentz}d). We find empirically that the symmetric angular dependence can be almost entirely fitted by the anti-damping torque ($h_{\mathrm{z}} \propto \cos\theta$) and that the antisymmetric angular dependence can be almost entirely fitted by a field-like term ($h_{\mathrm{y}}$ independent of angle). Small additional terms which are not consistent in size or sign from device to device are needed for the fitting ($h_{\mathrm{z}}$ independent of angle and $h_{\mathrm{y}} \propto \cos\theta$). These terms are consistent with additional field-like and anti-damping torques with symmetry $\boldsymbol{\tau} \propto \mathbf{\hat{z}\times\hat{m}}$ and $\boldsymbol{\tau} \propto \mathbf{\hat{m}\times\boldsymbol{\hat{z}}\times\hat{m}}$ respectively. Most significantly, we see that as the cobalt thickness is reduced from 3 to 1 nm, the sign of the symmetric voltage stays constant, whilst the sign of the antisymmetric voltage flips (see figures \ref{im:lorentz}b and c). This indicates that as the cobalt thickness is reduced the direction of the field-like torque reverses. The voltages measured scale linearly with power (figure \ref{im:lorentz}e), showing the torques are proportional to current density (as $V_{\mathrm{mix}}$ is proportional to microwave current).

\begin{figure}
\centering
\includegraphics[]{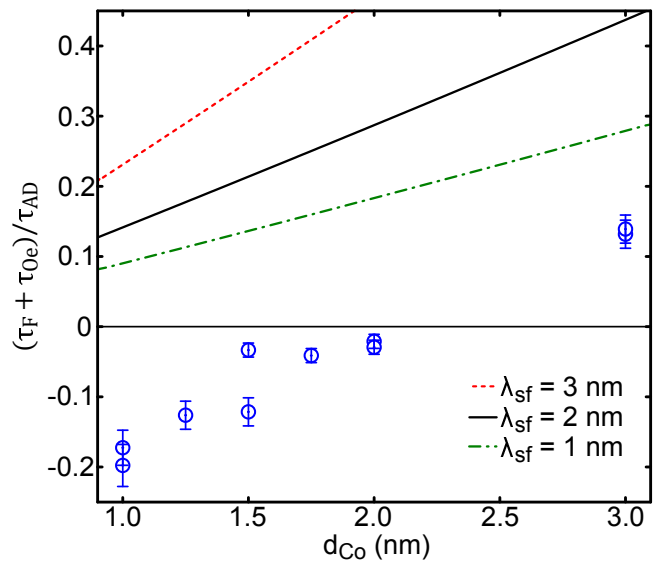}
\caption{The ratio of the total field-like torque to anti-damping torque is calculated for a series of cobalt thicknesses ($d_{\mathrm{Pt}}=3$ nm). The error bars show the uncertainty due to the fitting of the torque values to the data. Further scatter in the data may be due to variation within each layer studied. Additionally we also show the calculated torque ratio for the model presented by Liu \textit{et al.} \cite{Liu2011} where the field-like torque is purely Oersted and the anti-damping torque is purely due to the spin-transfer torque of the spin-Hall spin current. We plot the theoretical curves for $\theta_{\mathrm{SH}}=0.08$ (the same as Liu \textit{et al.} reported for their self-calibration method), for Pt spin-diffusion lengths of 1,2 and 3 nm. In thicker Co films the torque ratio begins to converge with this model with a spin-diffusion length of 1 nm, but the deviation, including a sign change, in thinner films shows an additional field-like torque becoming increasingly strong.}	
\label{im:fields}
\end{figure}

The microwave currents have previously been directly calibrated in similar measurements using a bolometric technique.\cite{fang2011spin} In this experiment we cannot use this technique because of the small temperature coefficient of resistance in the samples. Instead, we compare the relative sizes of the fitted torques induced by the same current. Figure \ref{im:fields} shows the ratio of the total field-like to anti-damping torques ($(\tau_{\mathrm{F}}+\tau_{\mathrm{Oe}})/\tau_{\mathrm{AD}}$) for the range of cobalt thicknesses measured. We also show the calculated ratio for the case where the field-like torque is purely Oersted and the anti-damping torque is due to the spin-transfer torque of the spin-Hall spin-current. The calculated ratio depends on the values of $\theta_{\mathrm{SH}}$ and the spin-diffusion length, $\lambda_{\mathrm{sf}}$, of platinum.\cite{Liu2011} Here we have used $\theta_{\mathrm{SH}}=0.08$ (as reported by Liu \textit{et al.}\cite{Liu2011}) for $\lambda_{\mathrm{sf}}=$ 1,2 and 3 nm. For this calculation we also use saturation magnetisation values, found from SQUID measurements of these layers, of $\mu_0M_{\mathrm{S}}=1.45$ T. As the Co layer becomes thicker we find the ratio converges with the theoretical curve for $\lambda_{\mathrm{sf}}=1$ nm. However, as $d_{\mathrm{Co}}$ reduces below around 2 nm, the ratio become negative and diverges from the theoretical curves, indicating the presence of an additional field-like torque, $\tau_{\mathrm{F}}$, which increasingly opposes the Oersted torque. We note that if we use more conservative values for our theoretical modelling (larger $\lambda_{\mathrm{sf}}$, smaller $\theta_{\mathrm{SH}}$), $\tau_{\mathrm{F}}$ is even larger.

This reversal in sign of the total field-like torque has not been observed before in Co/Pt. We note that the sign of our $\tau_{\mathrm{F}}$ and $\tau_{\mathrm{AD}}$ is consistent with the torques observed by Garello \textit{et al.} at low frequency in an AlOx/Co(0.6 nm)/Pt(3nm) device.\cite{Garello2013} Equally, in the $d_{\mathrm{Co}}= 3$ nm layer, where $\tau_{\mathrm{F}}$ is weakest, the torques resemble those measured by Liu \textit{et al.} at microwave frequencies in Py(4 nm)/Pt(6 nm) \cite{Liu2011} and CoFeB(3 nm)/Pt(6 nm).\cite{Liu04052012}

Kim \textit{et al.} have studied the torques at low frequency as a function of ferromagnet thickness (0.9 to 1.3 nm) in CoFeB/Ta(1 nm).\cite{kim2012layer} They observed a constant $\tau_{\mathrm{AD}}$ with opposite sign, because the spin-Hall angle is negative in Ta. $\tau_{\mathrm{F}}$ increased in the thinner ferromagnet layers, but in contrast to our observation, added to the Oersted torque.

Fan \textit{et al.} have measured the torques at low frequency, with a Cu spacer layer inserted between a Py and Pt layer.\cite{fan2013observation} A field-like torque was observed even with the spacer layer, and reduced with increasing spacer thickness, indicating that the torque was likely to be a non-adiabatic STT. As the ferromagnet thickness was reduced, the torque increased and added to the Oersted torque. This is the opposite sign to the $\tau_{\mathrm{F}}$ we have observed in Co/Pt. Fan \textit{et al.} also studied CoFeB/Ta layers using electrically driven FMR. It could be seen that as the ferromagnet layer is reduced, the field-like torque increases, and opposes the Oersted field. This is the opposite sign to the observation of Kim \textit{et al.}


When trying to reconcile these previous measurements with our own, we consider it likely that differing material parameters in each experiment, the quality of the interfaces and the degree of oxidation of the additional ferromagnet interface could give quite different results. Nonetheless, the trend and sign of the field-like torque we observe is consistent with the studies by Liu \textit{et al.} and Garello \textit{et al.} Furthermore, if the direction of the Rashba field is inverted in CoFeB/Ta compared to Co/Pt,\cite{suzuki2011current} or if the non-adiabatic STT depends on the negative sign of the SHE in Ta, our results can also be consistent with \textit{Kim et al.} and an earlier measurement by Suzuki \textit{et al.}\cite{suzuki2011current} That the sign of $\tau_{\mathrm{F}}$ we measure opposes the Oersted field and is the opposite of the one measured by Fan \textit{et al.} in Py/Pt can be explained by the torques having different origins. The measurements by Fan \textit{et al.} with Cu spacers strongly indicate a non-adiabatic STT origin in their case. In contrast, our $\tau_{\mathrm{F}}$ opposing the Oersted field is consistent with a Rashba field, with opposite sign to the non-adiabatic STT, dominating in our material. 


In conclusion, using an electrically driven FMR technique, we have observed an increase in the field-like torque, as a proportion of the total spin-orbit torques, as the cobalt layer is reduced from 3 to 1 nm. This field-like torque opposes the Oersted torque. The enhancement in the torque is consistent with a Rashba field, and takes the opposite sign to previous measurements in Py/Pt where the torque was shown to mostly originate non-locally from the interface, in the platinum layer. Whilst this is consistent with a Rashba origin of the field-like torque, we cannot rule out a contribution from a non-adiabatic STT.

AWR acknowledges support from an EPSRC Career Acceleration Fellowship grant EP/H003487/1. A.J.F acknowledges support from the Hitachi research fellowship and a Royal society research grant (RG110616).


\begin{thebibliography}{30}%
\makeatletter
\providecommand \@ifxundefined [1]{%
 \@ifx{#1\undefined}
}%
\providecommand \@ifnum [1]{%
 \ifnum #1\expandafter \@firstoftwo
 \else \expandafter \@secondoftwo
 \fi
}%
\providecommand \@ifx [1]{%
 \ifx #1\expandafter \@firstoftwo
 \else \expandafter \@secondoftwo
 \fi
}%
\providecommand \natexlab [1]{#1}%
\providecommand \enquote  [1]{``#1''}%
\providecommand \bibnamefont  [1]{#1}%
\providecommand \bibfnamefont [1]{#1}%
\providecommand \citenamefont [1]{#1}%
\providecommand \href@noop [0]{\@secondoftwo}%
\providecommand \href [0]{\begingroup \@sanitize@url \@href}%
\providecommand \@href[1]{\@@startlink{#1}\@@href}%
\providecommand \@@href[1]{\endgroup#1\@@endlink}%
\providecommand \@sanitize@url [0]{\catcode `\\12\catcode `\$12\catcode
  `\&12\catcode `\#12\catcode `\^12\catcode `\_12\catcode `\%12\relax}%
\providecommand \@@startlink[1]{}%
\providecommand \@@endlink[0]{}%
\providecommand \url  [0]{\begingroup\@sanitize@url \@url }%
\providecommand \@url [1]{\endgroup\@href {#1}{\urlprefix }}%
\providecommand \urlprefix  [0]{URL }%
\providecommand \Eprint [0]{\href }%
\providecommand \doibase [0]{http://dx.doi.org/}%
\providecommand \selectlanguage [0]{\@gobble}%
\providecommand \bibinfo  [0]{\@secondoftwo}%
\providecommand \bibfield  [0]{\@secondoftwo}%
\providecommand \translation [1]{[#1]}%
\providecommand \BibitemOpen [0]{}%
\providecommand \bibitemStop [0]{}%
\providecommand \bibitemNoStop [0]{.\EOS\space}%
\providecommand \EOS [0]{\spacefactor3000\relax}%
\providecommand \BibitemShut  [1]{\csname bibitem#1\endcsname}%
\let\auto@bib@innerbib\@empty
\bibitem [{\citenamefont {Hirsch}(1999)}]{hirsch1999spin}%
  \BibitemOpen
  \bibfield  {author} {\bibinfo {author} {\bibfnamefont {J.}~\bibnamefont
  {Hirsch}},\ }\href@noop {} {\bibfield  {journal} {\bibinfo  {journal} {Phys.
  Rev. Lett.}\ }\textbf {\bibinfo {volume} {83}},\ \bibinfo {pages} {1834}
  (\bibinfo {year} {1999})}\BibitemShut {NoStop}%
\bibitem [{\citenamefont {Sinova}\ \emph {et~al.}(2004)\citenamefont {Sinova},
  \citenamefont {Culcer}, \citenamefont {Niu}, \citenamefont {Sinitsyn},
  \citenamefont {Jungwirth},\ and\ \citenamefont
  {MacDonald}}]{sinova2004universal}%
  \BibitemOpen
  \bibfield  {author} {\bibinfo {author} {\bibfnamefont {J.}~\bibnamefont
  {Sinova}}, \bibinfo {author} {\bibfnamefont {D.}~\bibnamefont {Culcer}},
  \bibinfo {author} {\bibfnamefont {Q.}~\bibnamefont {Niu}}, \bibinfo {author}
  {\bibfnamefont {N.}~\bibnamefont {Sinitsyn}}, \bibinfo {author}
  {\bibfnamefont {T.}~\bibnamefont {Jungwirth}}, \ and\ \bibinfo {author}
  {\bibfnamefont {A.}~\bibnamefont {MacDonald}},\ }\href@noop {} {\bibfield
  {journal} {\bibinfo  {journal} {Phys. Rev. Lett.}\ }\textbf {\bibinfo
  {volume} {92}},\ \bibinfo {pages} {126603} (\bibinfo {year}
  {2004})}\BibitemShut {NoStop}%
\bibitem [{\citenamefont {Kato}\ \emph {et~al.}(2004)\citenamefont {Kato},
  \citenamefont {Myers}, \citenamefont {Gossard},\ and\ \citenamefont
  {Awschalom}}]{kato2004observation}%
  \BibitemOpen
  \bibfield  {author} {\bibinfo {author} {\bibfnamefont {Y.}~\bibnamefont
  {Kato}}, \bibinfo {author} {\bibfnamefont {R.}~\bibnamefont {Myers}},
  \bibinfo {author} {\bibfnamefont {A.}~\bibnamefont {Gossard}}, \ and\
  \bibinfo {author} {\bibfnamefont {D.}~\bibnamefont {Awschalom}},\ }\href@noop
  {} {\bibfield  {journal} {\bibinfo  {journal} {Science}\ }\textbf {\bibinfo
  {volume} {306}},\ \bibinfo {pages} {1910} (\bibinfo {year}
  {2004})}\BibitemShut {NoStop}%
\bibitem [{\citenamefont {Wunderlich}\ \emph {et~al.}(2005)\citenamefont
  {Wunderlich}, \citenamefont {Kaestner}, \citenamefont {Sinova},\ and\
  \citenamefont {Jungwirth}}]{wunderlich2005experimental}%
  \BibitemOpen
  \bibfield  {author} {\bibinfo {author} {\bibfnamefont {J.}~\bibnamefont
  {Wunderlich}}, \bibinfo {author} {\bibfnamefont {B.}~\bibnamefont
  {Kaestner}}, \bibinfo {author} {\bibfnamefont {J.}~\bibnamefont {Sinova}}, \
  and\ \bibinfo {author} {\bibfnamefont {T.}~\bibnamefont {Jungwirth}},\
  }\href@noop {} {\bibfield  {journal} {\bibinfo  {journal} {Phys. Rev. Lett.}\
  }\textbf {\bibinfo {volume} {94}},\ \bibinfo {pages} {047204} (\bibinfo
  {year} {2005})}\BibitemShut {NoStop}%
\bibitem [{\citenamefont {Liu}\ \emph {et~al.}(2011)\citenamefont {Liu},
  \citenamefont {Moriyama}, \citenamefont {Ralph},\ and\ \citenamefont
  {Buhrman}}]{Liu2011}%
  \BibitemOpen
  \bibfield  {author} {\bibinfo {author} {\bibfnamefont {L.}~\bibnamefont
  {Liu}}, \bibinfo {author} {\bibfnamefont {T.}~\bibnamefont {Moriyama}},
  \bibinfo {author} {\bibfnamefont {D.~C.}\ \bibnamefont {Ralph}}, \ and\
  \bibinfo {author} {\bibfnamefont {R.~A.}\ \bibnamefont {Buhrman}},\ }\href
  {\doibase 10.1103/PhysRevLett.106.036601} {\bibfield  {journal} {\bibinfo
  {journal} {Phys. Rev. Lett.}\ }\textbf {\bibinfo {volume} {106}},\ \bibinfo
  {pages} {036601} (\bibinfo {year} {2011})}\BibitemShut {NoStop}%
\bibitem [{\citenamefont {Liu}\ \emph {et~al.}(2012{\natexlab{a}})\citenamefont
  {Liu}, \citenamefont {Pai}, \citenamefont {Li}, \citenamefont {Tseng},
  \citenamefont {Ralph},\ and\ \citenamefont {Buhrman}}]{Liu04052012}%
  \BibitemOpen
  \bibfield  {author} {\bibinfo {author} {\bibfnamefont {L.}~\bibnamefont
  {Liu}}, \bibinfo {author} {\bibfnamefont {C.-F.}\ \bibnamefont {Pai}},
  \bibinfo {author} {\bibfnamefont {Y.}~\bibnamefont {Li}}, \bibinfo {author}
  {\bibfnamefont {H.~W.}\ \bibnamefont {Tseng}}, \bibinfo {author}
  {\bibfnamefont {D.~C.}\ \bibnamefont {Ralph}}, \ and\ \bibinfo {author}
  {\bibfnamefont {R.~A.}\ \bibnamefont {Buhrman}},\ }\href {\doibase
  10.1126/science.1218197} {\bibfield  {journal} {\bibinfo  {journal}
  {Science}\ }\textbf {\bibinfo {volume} {336}},\ \bibinfo {pages} {555}
  (\bibinfo {year} {2012}{\natexlab{a}})}\BibitemShut {NoStop}%
\bibitem [{\citenamefont {Liu}\ \emph {et~al.}(2012{\natexlab{b}})\citenamefont
  {Liu}, \citenamefont {Lee}, \citenamefont {Gudmundsen}, \citenamefont
  {Ralph},\ and\ \citenamefont {Buhrman}}]{Liu2012current}%
  \BibitemOpen
  \bibfield  {author} {\bibinfo {author} {\bibfnamefont {L.}~\bibnamefont
  {Liu}}, \bibinfo {author} {\bibfnamefont {O.~J.}\ \bibnamefont {Lee}},
  \bibinfo {author} {\bibfnamefont {T.~J.}\ \bibnamefont {Gudmundsen}},
  \bibinfo {author} {\bibfnamefont {D.~C.}\ \bibnamefont {Ralph}}, \ and\
  \bibinfo {author} {\bibfnamefont {R.~A.}\ \bibnamefont {Buhrman}},\ }\href
  {\doibase 10.1103/PhysRevLett.109.096602} {\bibfield  {journal} {\bibinfo
  {journal} {Phys. Rev. Lett.}\ }\textbf {\bibinfo {volume} {109}},\ \bibinfo
  {pages} {096602} (\bibinfo {year} {2012}{\natexlab{b}})}\BibitemShut
  {NoStop}%
\bibitem [{\citenamefont {Slonczewski}(1996)}]{slonczewski1996current}%
  \BibitemOpen
  \bibfield  {author} {\bibinfo {author} {\bibfnamefont {J.}~\bibnamefont
  {Slonczewski}},\ }\href@noop {} {\bibfield  {journal} {\bibinfo  {journal}
  {J. Magn. Magn. Mater.}\ }\textbf {\bibinfo {volume} {159}},\ \bibinfo
  {pages} {L1} (\bibinfo {year} {1996})}\BibitemShut {NoStop}%
\bibitem [{\citenamefont {Berger}(1996)}]{berger1996emission}%
  \BibitemOpen
  \bibfield  {author} {\bibinfo {author} {\bibfnamefont {L.}~\bibnamefont
  {Berger}},\ }\href {\doibase 10.1103/PhysRevB.54.9353} {\bibfield  {journal}
  {\bibinfo  {journal} {Phys. Rev. B}\ }\textbf {\bibinfo {volume} {54}},\
  \bibinfo {pages} {9353} (\bibinfo {year} {1996})}\BibitemShut {NoStop}%
\bibitem [{\citenamefont {Zhang}, \citenamefont {Levy},\ and\ \citenamefont
  {Fert}(2002)}]{Zhang2002}%
  \BibitemOpen
  \bibfield  {author} {\bibinfo {author} {\bibfnamefont {S.}~\bibnamefont
  {Zhang}}, \bibinfo {author} {\bibfnamefont {P.~M.}\ \bibnamefont {Levy}}, \
  and\ \bibinfo {author} {\bibfnamefont {A.}~\bibnamefont {Fert}},\ }\href
  {\doibase 10.1103/PhysRevLett.88.236601} {\bibfield  {journal} {\bibinfo
  {journal} {Phys. Rev. Lett.}\ }\textbf {\bibinfo {volume} {88}},\ \bibinfo
  {pages} {236601} (\bibinfo {year} {2002})}\BibitemShut {NoStop}%
\bibitem [{\citenamefont {Zimmler}\ \emph {et~al.}(2004)\citenamefont
  {Zimmler}, \citenamefont {\"Ozyilmaz}, \citenamefont {Chen}, \citenamefont
  {Kent}, \citenamefont {Sun}, \citenamefont {Rooks},\ and\ \citenamefont
  {Koch}}]{zimmler2004}%
  \BibitemOpen
  \bibfield  {author} {\bibinfo {author} {\bibfnamefont {M.~A.}\ \bibnamefont
  {Zimmler}}, \bibinfo {author} {\bibfnamefont {B.}~\bibnamefont {\"Ozyilmaz}},
  \bibinfo {author} {\bibfnamefont {W.}~\bibnamefont {Chen}}, \bibinfo {author}
  {\bibfnamefont {A.~D.}\ \bibnamefont {Kent}}, \bibinfo {author}
  {\bibfnamefont {J.~Z.}\ \bibnamefont {Sun}}, \bibinfo {author} {\bibfnamefont
  {M.~J.}\ \bibnamefont {Rooks}}, \ and\ \bibinfo {author} {\bibfnamefont
  {R.~H.}\ \bibnamefont {Koch}},\ }\href {\doibase 10.1103/PhysRevB.70.184438}
  {\bibfield  {journal} {\bibinfo  {journal} {Phys. Rev. B}\ }\textbf {\bibinfo
  {volume} {70}},\ \bibinfo {pages} {184438} (\bibinfo {year}
  {2004})}\BibitemShut {NoStop}%
\bibitem [{\citenamefont {Sankey}\ \emph {et~al.}(2007)\citenamefont {Sankey},
  \citenamefont {Cui}, \citenamefont {Sun}, \citenamefont {Slonczewski},
  \citenamefont {Buhrman},\ and\ \citenamefont
  {Ralph}}]{sankey2007measurement}%
  \BibitemOpen
  \bibfield  {author} {\bibinfo {author} {\bibfnamefont {J.~C.}\ \bibnamefont
  {Sankey}}, \bibinfo {author} {\bibfnamefont {Y.-T.}\ \bibnamefont {Cui}},
  \bibinfo {author} {\bibfnamefont {J.~Z.}\ \bibnamefont {Sun}}, \bibinfo
  {author} {\bibfnamefont {J.~C.}\ \bibnamefont {Slonczewski}}, \bibinfo
  {author} {\bibfnamefont {R.~A.}\ \bibnamefont {Buhrman}}, \ and\ \bibinfo
  {author} {\bibfnamefont {D.~C.}\ \bibnamefont {Ralph}},\ }\href@noop {}
  {\bibfield  {journal} {\bibinfo  {journal} {Nature Phys.}\ }\textbf {\bibinfo
  {volume} {4}},\ \bibinfo {pages} {67} (\bibinfo {year} {2007})}\BibitemShut
  {NoStop}%
\bibitem [{\citenamefont {Manchon}\ and\ \citenamefont
  {Zhang}(2008)}]{Manchon2008}%
  \BibitemOpen
  \bibfield  {author} {\bibinfo {author} {\bibfnamefont {A.}~\bibnamefont
  {Manchon}}\ and\ \bibinfo {author} {\bibfnamefont {S.}~\bibnamefont
  {Zhang}},\ }\href {\doibase 10.1103/PhysRevB.78.212405} {\bibfield  {journal}
  {\bibinfo  {journal} {Phys. Rev. B}\ }\textbf {\bibinfo {volume} {78}},\
  \bibinfo {pages} {212405} (\bibinfo {year} {2008})}\BibitemShut {NoStop}%
\bibitem [{\citenamefont {Obata}\ and\ \citenamefont
  {Tatara}(2008)}]{Obata2008}%
  \BibitemOpen
  \bibfield  {author} {\bibinfo {author} {\bibfnamefont {K.}~\bibnamefont
  {Obata}}\ and\ \bibinfo {author} {\bibfnamefont {G.}~\bibnamefont {Tatara}},\
  }\href {\doibase 10.1103/PhysRevB.77.214429} {\bibfield  {journal} {\bibinfo
  {journal} {Phys. Rev. B}\ }\textbf {\bibinfo {volume} {77}},\ \bibinfo
  {pages} {214429} (\bibinfo {year} {2008})}\BibitemShut {NoStop}%
\bibitem [{\citenamefont {Manchon}\ and\ \citenamefont
  {Zhang}(2009)}]{Manchon2009}%
  \BibitemOpen
  \bibfield  {author} {\bibinfo {author} {\bibfnamefont {A.}~\bibnamefont
  {Manchon}}\ and\ \bibinfo {author} {\bibfnamefont {S.}~\bibnamefont
  {Zhang}},\ }\href {\doibase 10.1103/PhysRevB.79.094422} {\bibfield  {journal}
  {\bibinfo  {journal} {Phys. Rev. B}\ }\textbf {\bibinfo {volume} {79}},\
  \bibinfo {pages} {094422} (\bibinfo {year} {2009})}\BibitemShut {NoStop}%
\bibitem [{\citenamefont {Miron}\ \emph {et~al.}(2010)\citenamefont {Miron},
  \citenamefont {Gaudin}, \citenamefont {Auffret}, \citenamefont {Rodmacq},
  \citenamefont {Schuhl}, \citenamefont {Pizzini}, \citenamefont {Vogel},\ and\
  \citenamefont {Gambardella}}]{miron2010current}%
  \BibitemOpen
  \bibfield  {author} {\bibinfo {author} {\bibfnamefont {I.~M.}\ \bibnamefont
  {Miron}}, \bibinfo {author} {\bibfnamefont {G.}~\bibnamefont {Gaudin}},
  \bibinfo {author} {\bibfnamefont {S.}~\bibnamefont {Auffret}}, \bibinfo
  {author} {\bibfnamefont {B.}~\bibnamefont {Rodmacq}}, \bibinfo {author}
  {\bibfnamefont {A.}~\bibnamefont {Schuhl}}, \bibinfo {author} {\bibfnamefont
  {S.}~\bibnamefont {Pizzini}}, \bibinfo {author} {\bibfnamefont
  {J.}~\bibnamefont {Vogel}}, \ and\ \bibinfo {author} {\bibfnamefont
  {P.}~\bibnamefont {Gambardella}},\ }\href@noop {} {\bibfield  {journal}
  {\bibinfo  {journal} {Nat. Mater.}\ }\textbf {\bibinfo {volume} {9}},\
  \bibinfo {pages} {230} (\bibinfo {year} {2010})}\BibitemShut {NoStop}%
\bibitem [{\citenamefont {Miron}\ \emph
  {et~al.}(2011{\natexlab{a}})\citenamefont {Miron}, \citenamefont {Moore},
  \citenamefont {Szambolics}, \citenamefont {Buda-Prejbeanu}, \citenamefont
  {Auffret}, \citenamefont {Rodmacq}, \citenamefont {Pizzini}, \citenamefont
  {Vogel}, \citenamefont {Bonfim}, \citenamefont {Schuhl} \emph
  {et~al.}}]{miron2011fast}%
  \BibitemOpen
  \bibfield  {author} {\bibinfo {author} {\bibfnamefont {I.~M.}\ \bibnamefont
  {Miron}}, \bibinfo {author} {\bibfnamefont {T.}~\bibnamefont {Moore}},
  \bibinfo {author} {\bibfnamefont {H.}~\bibnamefont {Szambolics}}, \bibinfo
  {author} {\bibfnamefont {L.~D.}\ \bibnamefont {Buda-Prejbeanu}}, \bibinfo
  {author} {\bibfnamefont {S.}~\bibnamefont {Auffret}}, \bibinfo {author}
  {\bibfnamefont {B.}~\bibnamefont {Rodmacq}}, \bibinfo {author} {\bibfnamefont
  {S.}~\bibnamefont {Pizzini}}, \bibinfo {author} {\bibfnamefont
  {J.}~\bibnamefont {Vogel}}, \bibinfo {author} {\bibfnamefont
  {M.}~\bibnamefont {Bonfim}}, \bibinfo {author} {\bibfnamefont
  {A.}~\bibnamefont {Schuhl}},  \emph {et~al.},\ }\href@noop {} {\bibfield
  {journal} {\bibinfo  {journal} {Nat. Mater.}\ }\textbf {\bibinfo {volume}
  {10}},\ \bibinfo {pages} {419} (\bibinfo {year}
  {2011}{\natexlab{a}})}\BibitemShut {NoStop}%
\bibitem [{\citenamefont {Pi}\ \emph {et~al.}(2010)\citenamefont {Pi},
  \citenamefont {Won~Kim}, \citenamefont {Bae}, \citenamefont {Lee},
  \citenamefont {Cho}, \citenamefont {Kim},\ and\ \citenamefont
  {Seo}}]{pi2010tilting}%
  \BibitemOpen
  \bibfield  {author} {\bibinfo {author} {\bibfnamefont {U.~H.}\ \bibnamefont
  {Pi}}, \bibinfo {author} {\bibfnamefont {K.}~\bibnamefont {Won~Kim}},
  \bibinfo {author} {\bibfnamefont {J.~Y.}\ \bibnamefont {Bae}}, \bibinfo
  {author} {\bibfnamefont {S.~C.}\ \bibnamefont {Lee}}, \bibinfo {author}
  {\bibfnamefont {Y.~J.}\ \bibnamefont {Cho}}, \bibinfo {author} {\bibfnamefont
  {K.~S.}\ \bibnamefont {Kim}}, \ and\ \bibinfo {author} {\bibfnamefont
  {S.}~\bibnamefont {Seo}},\ }\href {\doibase
  http://dx.doi.org/10.1063/1.3502596} {\bibfield  {journal} {\bibinfo
  {journal} {Appl. Phys. Lett.}\ }\textbf {\bibinfo {volume} {97}},\ \bibinfo
  {eid} {162507} (\bibinfo {year} {2010})}\BibitemShut {NoStop}%
\bibitem [{\citenamefont {Suzuki}\ \emph {et~al.}(2011)\citenamefont {Suzuki},
  \citenamefont {Fukami}, \citenamefont {Ishiwata}, \citenamefont {Yamanouchi},
  \citenamefont {Ikeda}, \citenamefont {Kasai},\ and\ \citenamefont
  {Ohno}}]{suzuki2011current}%
  \BibitemOpen
  \bibfield  {author} {\bibinfo {author} {\bibfnamefont {T.}~\bibnamefont
  {Suzuki}}, \bibinfo {author} {\bibfnamefont {S.}~\bibnamefont {Fukami}},
  \bibinfo {author} {\bibfnamefont {N.}~\bibnamefont {Ishiwata}}, \bibinfo
  {author} {\bibfnamefont {M.}~\bibnamefont {Yamanouchi}}, \bibinfo {author}
  {\bibfnamefont {S.}~\bibnamefont {Ikeda}}, \bibinfo {author} {\bibfnamefont
  {N.}~\bibnamefont {Kasai}}, \ and\ \bibinfo {author} {\bibfnamefont
  {H.}~\bibnamefont {Ohno}},\ }\href {\doibase
  http://dx.doi.org/10.1063/1.3579155} {\bibfield  {journal} {\bibinfo
  {journal} {Appl. Phys. Lett.}\ }\textbf {\bibinfo {volume} {98}},\ \bibinfo
  {eid} {142505} (\bibinfo {year} {2011})}\BibitemShut {NoStop}%
\bibitem [{\citenamefont {Miron}\ \emph
  {et~al.}(2011{\natexlab{b}})\citenamefont {Miron}, \citenamefont {Garello},
  \citenamefont {Gaudin}, \citenamefont {Zermatten}, \citenamefont {Costache},
  \citenamefont {Auffret}, \citenamefont {Bandiera}, \citenamefont {Rodmacq},
  \citenamefont {Schuhl},\ and\ \citenamefont
  {Gambardella}}]{miron2011perpendicular}%
  \BibitemOpen
  \bibfield  {author} {\bibinfo {author} {\bibfnamefont {I.~M.}\ \bibnamefont
  {Miron}}, \bibinfo {author} {\bibfnamefont {K.}~\bibnamefont {Garello}},
  \bibinfo {author} {\bibfnamefont {G.}~\bibnamefont {Gaudin}}, \bibinfo
  {author} {\bibfnamefont {P.-J.}\ \bibnamefont {Zermatten}}, \bibinfo {author}
  {\bibfnamefont {M.~V.}\ \bibnamefont {Costache}}, \bibinfo {author}
  {\bibfnamefont {S.}~\bibnamefont {Auffret}}, \bibinfo {author} {\bibfnamefont
  {S.}~\bibnamefont {Bandiera}}, \bibinfo {author} {\bibfnamefont
  {B.}~\bibnamefont {Rodmacq}}, \bibinfo {author} {\bibfnamefont
  {A.}~\bibnamefont {Schuhl}}, \ and\ \bibinfo {author} {\bibfnamefont
  {P.}~\bibnamefont {Gambardella}},\ }\href@noop {} {\bibfield  {journal}
  {\bibinfo  {journal} {Nature}\ }\textbf {\bibinfo {volume} {476}},\ \bibinfo
  {pages} {189} (\bibinfo {year} {2011}{\natexlab{b}})}\BibitemShut {NoStop}%
\bibitem [{\citenamefont {Garello}\ \emph {et~al.}(2013)\citenamefont
  {Garello}, \citenamefont {Miron}, \citenamefont {Avci}, \citenamefont
  {Freimuth}, \citenamefont {Mokrousov}, \citenamefont {Blugel}, \citenamefont
  {Boulle}, \citenamefont {Gaudin},\ and\ \citenamefont
  {Gambardella}}]{Garello2013}%
  \BibitemOpen
  \bibfield  {author} {\bibinfo {author} {\bibfnamefont {K.}~\bibnamefont
  {Garello}}, \bibinfo {author} {\bibfnamefont {I.~M.}\ \bibnamefont {Miron}},
  \bibinfo {author} {\bibfnamefont {C.~O.}\ \bibnamefont {Avci}}, \bibinfo
  {author} {\bibfnamefont {F.}~\bibnamefont {Freimuth}}, \bibinfo {author}
  {\bibfnamefont {Y.}~\bibnamefont {Mokrousov}}, \bibinfo {author}
  {\bibfnamefont {S.}~\bibnamefont {Blugel}, \bibfnamefont
  {Stefanand~Auffret}}, \bibinfo {author} {\bibfnamefont {O.}~\bibnamefont
  {Boulle}}, \bibinfo {author} {\bibfnamefont {G.}~\bibnamefont {Gaudin}}, \
  and\ \bibinfo {author} {\bibfnamefont {P.}~\bibnamefont {Gambardella}},\
  }\href {\doibase 10.1038/nnano.2013.145} {\bibfield  {journal} {\bibinfo
  {journal} {Nat. Nanotechnol.}\ }\textbf {\bibinfo {volume} {8}},\ \bibinfo
  {pages} {587} (\bibinfo {year} {2013})}\BibitemShut {NoStop}%
\bibitem [{\citenamefont {Kurebayashi}\ \emph {et~al.}(2013)\citenamefont
  {Kurebayashi}, \citenamefont {Sinova}, \citenamefont {Fang}, \citenamefont
  {Irvine}, \citenamefont {Wunderlich}, \citenamefont {Novak}, \citenamefont
  {Campion}, \citenamefont {Gallagher}, \citenamefont {Vehstedt}, \citenamefont
  {Zarbo} \emph {et~al.}}]{kurebayashi2013observation}%
  \BibitemOpen
  \bibfield  {author} {\bibinfo {author} {\bibfnamefont {H.}~\bibnamefont
  {Kurebayashi}}, \bibinfo {author} {\bibfnamefont {J.}~\bibnamefont {Sinova}},
  \bibinfo {author} {\bibfnamefont {D.}~\bibnamefont {Fang}}, \bibinfo {author}
  {\bibfnamefont {A.}~\bibnamefont {Irvine}}, \bibinfo {author} {\bibfnamefont
  {J.}~\bibnamefont {Wunderlich}}, \bibinfo {author} {\bibfnamefont
  {V.}~\bibnamefont {Novak}}, \bibinfo {author} {\bibfnamefont
  {R.}~\bibnamefont {Campion}}, \bibinfo {author} {\bibfnamefont
  {B.}~\bibnamefont {Gallagher}}, \bibinfo {author} {\bibfnamefont
  {E.}~\bibnamefont {Vehstedt}}, \bibinfo {author} {\bibfnamefont
  {L.}~\bibnamefont {Zarbo}},  \emph {et~al.},\ }\href@noop {} {\bibfield
  {journal} {\bibinfo  {journal} {arXiv preprint arXiv:1306.1893}\ } (\bibinfo
  {year} {2013})}\BibitemShut {NoStop}%
\bibitem [{\citenamefont {Pesin}\ and\ \citenamefont
  {MacDonald}(2012)}]{Pesin2012}%
  \BibitemOpen
  \bibfield  {author} {\bibinfo {author} {\bibfnamefont {D.~A.}\ \bibnamefont
  {Pesin}}\ and\ \bibinfo {author} {\bibfnamefont {A.~H.}\ \bibnamefont
  {MacDonald}},\ }\href {\doibase 10.1103/PhysRevB.86.014416} {\bibfield
  {journal} {\bibinfo  {journal} {Phys. Rev. B}\ }\textbf {\bibinfo {volume}
  {86}},\ \bibinfo {pages} {014416} (\bibinfo {year} {2012})}\BibitemShut
  {NoStop}%
\bibitem [{\citenamefont {Wang}\ and\ \citenamefont
  {Manchon}(2012)}]{Wang2012}%
  \BibitemOpen
  \bibfield  {author} {\bibinfo {author} {\bibfnamefont {X.}~\bibnamefont
  {Wang}}\ and\ \bibinfo {author} {\bibfnamefont {A.}~\bibnamefont {Manchon}},\
  }\href {\doibase 10.1103/PhysRevLett.108.117201} {\bibfield  {journal}
  {\bibinfo  {journal} {Phys. Rev. Lett.}\ }\textbf {\bibinfo {volume} {108}},\
  \bibinfo {pages} {117201} (\bibinfo {year} {2012})}\BibitemShut {NoStop}%
\bibitem [{\citenamefont {Kim}\ \emph {et~al.}(2012)\citenamefont {Kim},
  \citenamefont {Sinha}, \citenamefont {Hayashi}, \citenamefont {Yamanouchi},
  \citenamefont {Fukami}, \citenamefont {Suzuki}, \citenamefont {Mitani},\ and\
  \citenamefont {Ohno}}]{kim2012layer}%
  \BibitemOpen
  \bibfield  {author} {\bibinfo {author} {\bibfnamefont {J.}~\bibnamefont
  {Kim}}, \bibinfo {author} {\bibfnamefont {J.}~\bibnamefont {Sinha}}, \bibinfo
  {author} {\bibfnamefont {M.}~\bibnamefont {Hayashi}}, \bibinfo {author}
  {\bibfnamefont {M.}~\bibnamefont {Yamanouchi}}, \bibinfo {author}
  {\bibfnamefont {S.}~\bibnamefont {Fukami}}, \bibinfo {author} {\bibfnamefont
  {T.}~\bibnamefont {Suzuki}}, \bibinfo {author} {\bibfnamefont
  {S.}~\bibnamefont {Mitani}}, \ and\ \bibinfo {author} {\bibfnamefont
  {H.}~\bibnamefont {Ohno}},\ }\href@noop {} {\bibfield  {journal} {\bibinfo
  {journal} {Nat. Mater.}\ } (\bibinfo {year} {2012})}\BibitemShut {NoStop}%
\bibitem [{\citenamefont {Fan}\ \emph {et~al.}(2013)\citenamefont {Fan},
  \citenamefont {Wu}, \citenamefont {Chen}, \citenamefont {Jerry},
  \citenamefont {Zhang},\ and\ \citenamefont {Xiao}}]{fan2013observation}%
  \BibitemOpen
  \bibfield  {author} {\bibinfo {author} {\bibfnamefont {X.}~\bibnamefont
  {Fan}}, \bibinfo {author} {\bibfnamefont {J.}~\bibnamefont {Wu}}, \bibinfo
  {author} {\bibfnamefont {Y.}~\bibnamefont {Chen}}, \bibinfo {author}
  {\bibfnamefont {M.~J.}\ \bibnamefont {Jerry}}, \bibinfo {author}
  {\bibfnamefont {H.}~\bibnamefont {Zhang}}, \ and\ \bibinfo {author}
  {\bibfnamefont {J.~Q.}\ \bibnamefont {Xiao}},\ }\href@noop {} {\bibfield
  {journal} {\bibinfo  {journal} {Nat. Commun.}\ }\textbf {\bibinfo {volume}
  {4}},\ \bibinfo {pages} {1799} (\bibinfo {year} {2013})}\BibitemShut
  {NoStop}%
\bibitem [{\citenamefont {Fang}\ \emph {et~al.}(2011)\citenamefont {Fang},
  \citenamefont {Kurebayashi}, \citenamefont {Wunderlich}, \citenamefont
  {V{\`y}born{\`y}}, \citenamefont {Z{\^a}rbo}, \citenamefont {Campion},
  \citenamefont {Casiraghi}, \citenamefont {Gallagher}, \citenamefont
  {Jungwirth},\ and\ \citenamefont {Ferguson}}]{fang2011spin}%
  \BibitemOpen
  \bibfield  {author} {\bibinfo {author} {\bibfnamefont {D.}~\bibnamefont
  {Fang}}, \bibinfo {author} {\bibfnamefont {H.}~\bibnamefont {Kurebayashi}},
  \bibinfo {author} {\bibfnamefont {J.}~\bibnamefont {Wunderlich}}, \bibinfo
  {author} {\bibfnamefont {K.}~\bibnamefont {V{\`y}born{\`y}}}, \bibinfo
  {author} {\bibfnamefont {L.}~\bibnamefont {Z{\^a}rbo}}, \bibinfo {author}
  {\bibfnamefont {R.}~\bibnamefont {Campion}}, \bibinfo {author} {\bibfnamefont
  {A.}~\bibnamefont {Casiraghi}}, \bibinfo {author} {\bibfnamefont
  {B.}~\bibnamefont {Gallagher}}, \bibinfo {author} {\bibfnamefont
  {T.}~\bibnamefont {Jungwirth}}, \ and\ \bibinfo {author} {\bibfnamefont
  {A.}~\bibnamefont {Ferguson}},\ }\href@noop {} {\bibfield  {journal}
  {\bibinfo  {journal} {Nat. Nanotechnol.}\ }\textbf {\bibinfo {volume} {6}},\
  \bibinfo {pages} {413} (\bibinfo {year} {2011})}\BibitemShut {NoStop}%
\bibitem [{\citenamefont {Fang}\ \emph {et~al.}(2012)\citenamefont {Fang},
  \citenamefont {Skinner}, \citenamefont {Kurebayashi}, \citenamefont
  {Campion}, \citenamefont {Gallagher},\ and\ \citenamefont
  {Ferguson}}]{fang2012electrical}%
  \BibitemOpen
  \bibfield  {author} {\bibinfo {author} {\bibfnamefont {D.}~\bibnamefont
  {Fang}}, \bibinfo {author} {\bibfnamefont {T.}~\bibnamefont {Skinner}},
  \bibinfo {author} {\bibfnamefont {H.}~\bibnamefont {Kurebayashi}}, \bibinfo
  {author} {\bibfnamefont {R.}~\bibnamefont {Campion}}, \bibinfo {author}
  {\bibfnamefont {B.}~\bibnamefont {Gallagher}}, \ and\ \bibinfo {author}
  {\bibfnamefont {A.}~\bibnamefont {Ferguson}},\ }\href@noop {} {\bibfield
  {journal} {\bibinfo  {journal} {Appl. Phys. Lett.}\ }\textbf {\bibinfo
  {volume} {101}},\ \bibinfo {pages} {182402} (\bibinfo {year}
  {2012})}\BibitemShut {NoStop}%
\bibitem [{\citenamefont {Ando}\ \emph {et~al.}(2011)\citenamefont {Ando},
  \citenamefont {Takahashi}, \citenamefont {Ieda}, \citenamefont {Kajiwara},
  \citenamefont {Nakayama}, \citenamefont {Yoshino}, \citenamefont {Harii},
  \citenamefont {Fujikawa}, \citenamefont {Matsuo}, \citenamefont {Maekawa},\
  and\ \citenamefont {Saitoh}}]{Ando2011}%
  \BibitemOpen
  \bibfield  {author} {\bibinfo {author} {\bibfnamefont {K.}~\bibnamefont
  {Ando}}, \bibinfo {author} {\bibfnamefont {S.}~\bibnamefont {Takahashi}},
  \bibinfo {author} {\bibfnamefont {J.}~\bibnamefont {Ieda}}, \bibinfo {author}
  {\bibfnamefont {Y.}~\bibnamefont {Kajiwara}}, \bibinfo {author}
  {\bibfnamefont {H.}~\bibnamefont {Nakayama}}, \bibinfo {author}
  {\bibfnamefont {T.}~\bibnamefont {Yoshino}}, \bibinfo {author} {\bibfnamefont
  {K.}~\bibnamefont {Harii}}, \bibinfo {author} {\bibfnamefont
  {Y.}~\bibnamefont {Fujikawa}}, \bibinfo {author} {\bibfnamefont
  {M.}~\bibnamefont {Matsuo}}, \bibinfo {author} {\bibfnamefont
  {S.}~\bibnamefont {Maekawa}}, \ and\ \bibinfo {author} {\bibfnamefont
  {E.}~\bibnamefont {Saitoh}},\ }\href {\doibase
  http://dx.doi.org/10.1063/1.3587173} {\bibfield  {journal} {\bibinfo
  {journal} {J. Appl. Phys.}\ }\textbf {\bibinfo {volume} {109}},\ \bibinfo
  {eid} {103913} (\bibinfo {year} {2011})}\BibitemShut {NoStop}%
\bibitem [{\citenamefont {Skinner}\ \emph {et~al.}(2013)\citenamefont
  {Skinner}, \citenamefont {Kurebayashi}, \citenamefont {Fang}, \citenamefont
  {Heiss}, \citenamefont {Irvine}, \citenamefont {Hindmarch}, \citenamefont
  {Wang}, \citenamefont {Rushforth},\ and\ \citenamefont
  {Ferguson}}]{skinner2013}%
  \BibitemOpen
  \bibfield  {author} {\bibinfo {author} {\bibfnamefont {T.~D.}\ \bibnamefont
  {Skinner}}, \bibinfo {author} {\bibfnamefont {H.}~\bibnamefont
  {Kurebayashi}}, \bibinfo {author} {\bibfnamefont {D.}~\bibnamefont {Fang}},
  \bibinfo {author} {\bibfnamefont {D.}~\bibnamefont {Heiss}}, \bibinfo
  {author} {\bibfnamefont {A.~C.}\ \bibnamefont {Irvine}}, \bibinfo {author}
  {\bibfnamefont {A.~T.}\ \bibnamefont {Hindmarch}}, \bibinfo {author}
  {\bibfnamefont {M.}~\bibnamefont {Wang}}, \bibinfo {author} {\bibfnamefont
  {A.~W.}\ \bibnamefont {Rushforth}}, \ and\ \bibinfo {author} {\bibfnamefont
  {A.~J.}\ \bibnamefont {Ferguson}},\ }\href {\doibase
  http://dx.doi.org/10.1063/1.4792693} {\bibfield  {journal} {\bibinfo
  {journal} {Appl. Phys. Lett.}\ }\textbf {\bibinfo {volume} {102}},\ \bibinfo
  {eid} {072401} (\bibinfo {year} {2013})}\BibitemShut {NoStop}%
\end{thebibliography}
\end{document}